\documentstyle[aps,12pt,epsf]{revtex}

\begin{document}

\title{The Skyrme Model $\pi NN$ Form Factor and the Sea Quark Distribution of the Nucleon}
\vskip 2cm
\author{R. J. Fries and A. Sch\"afer} 
\address{Institut f\"ur Theoretische Physik, Universit\"at Regensburg, \\
  D-93040 Regensburg, Germany} 
\maketitle 

\vskip 2cm

\begin{abstract}
We calculate the sea quark distribution of the nucleon in a meson cloud model.
The novel feature of our calculation is the implementation of a special 
$\pi NN$ form factor recently obtained by Holzwarth and Machleidt.
This form factor is hard for small and soft for large momentum transfers. 
We show that this feature leads to a substantial improvement.
\end{abstract}

\vskip 0.5cm
\noindent PACS: 13.75.Gx, 12.39.Dc, 13.60.Hb, 14.20.Dh

\vskip 0.5cm

Much effort has been invested over the years to calculate the contribution of 
mesons to the sea quark distributions in the nucleon at a scale of a few 
${\rm GeV}^2$ \cite{sull,awthomas,fra89,kumano,fra95}. 
In this contribution we do not want to comment on any of the existing models 
nor on their relative merits. 
We just want to demonstrate that the specific $\pi NN$ form factor, 
obtained recently by Holzwarth and Machleidt \cite{holzw96} for the Skyrme 
model, leads in a straightforward, lowest-order treatment of the problem to 
surprisingly good results.  
Let us note that the standard treatment, namely the convolution of the 
internal meson quark distribution with the momentum distribution of the mesons,
most notably the pion, has not necessarily to give the correct result as the 
pion is off-shell and the structure function of the off-shell pions should in 
principle be different from that of on-shell ones.
We shall keep all these caveats in mind.

The strength of meson cloud models in general for the description of sea 
quark distributions was demonstrated again lately by their relative ease in 
describing the observed asymmetry between $\bar{u}(x)$ and $\bar{d}(x)$ in the proton
\cite{hemi90,kumlond}.
Here the pions play a dominant role.
One crucial ingredient in the discussion is the choice of the form factor for 
the pion-nucleon coupling. In the past more or less involved parameterizations 
in monopole, dipole, or exponential form were used:
\begin{eqnarray}
  G_{mon}(t) &=& \frac{\Lambda_{mon}^2 - m_\pi^2}{\Lambda_{mon}^2 +t}, \\
  G_{di}(t) &=& {\left( \frac{\Lambda_{di}^2 - m_\pi^2}{\Lambda_{di}^2 +t} 
  \right)}^2, \\  
  G_{exp}(t) &=& \exp \left( -\frac{t+ m_\pi^2}{\Lambda_{exp}^2} \right).
\end{eqnarray}
Where $m_\pi$ is the pion mass, $t=-k^2$ is the negative four-momentum 
squared of the pion, and the $\Lambda$'s are cutoff parameters.

A general problem of this type of form factors is, that if, e.g.,\ the pionic 
contributions to the flavor symmetry breaking in the nucleon sea are fitted 
the resulting cutoffs are substantially smaller than those needed in nuclear 
physics, e.g.,\ for the Bonn potential \cite{bonn}. 
A careful discussion of this problem can be found in \cite{fra95}. 
In that paper Koepf, Frankfurt, and Strikman obtained $\Lambda_{exp}(\pi NN) 
= 850~{\rm MeV}$ from a fit of the pionic and kaonic contributions to 
$x (\bar{d}+\bar{u} -2\bar{s})$ with the same form factor and they obtained 
$\Lambda_{exp}(\pi NN) = 1000~{\rm MeV}$, when they allow three different 
cutoffs for the $\pi NN$, $\pi N\Delta$, and kaon contributions. 
This corresponds to quite soft monopole cutoffs of 660 MeV and 780 MeV --- 
according to the relation 
\begin{equation}
  \Lambda_{mon}\approx 0.62~\Lambda_{di} \approx 0.78~\Lambda_{exp}
\end{equation}
given by Kumano \cite{kumano} --- in comparison to 1.3 GeV often used for the 
Bonn potential. On the other hand, a monopole cutoff of 1.3 GeV would be too 
hard and lead to a large overestimation of the sea.

Therefore it is natural to search for alternative form factors.
As already shown in \cite{fra95} the dominant contribution to the pionic 
process comes from the region where $t$ is  above $20~m_\pi^2 \approx 0.4~{\rm GeV}^2$.
Smaller $t$ values are not very important. On the other hand the large cutoff 
parameters in nuclear physics are necessary to have a strong enough long-range 
part of the nuclear force, but they can lead to an insufficient suppression 
of high-t contributions \cite{holzw96}.
This implies that a form factor starting  at low t with a small slope and 
having a sudden and sufficiently rapid decrease at $t\approx 20~m_\pi^2$ 
could give us good results in both nuclear and deep-inelastic scattering (DIS) processes.       

It is therefore very interesting to note that a first principle calculation 
of the $\pi NN$ form factor within the Skyrme model \cite{holzw96} results 
in exactly such a type of form factor.
Holzwarth and Machleidt use the simple Skyrme Lagrangian
\begin{equation}
  {L} = \frac{f_\pi^2}{4} \int \left( -{\rm tr} L_\mu L^\mu + m^2 {\rm tr} 
   (U+U^\dagger -2)
   \right) d^3 x + \frac{1}{32e^2} \int {\rm tr} {[L_\mu,L_\nu]}^2 d^3 x
\end{equation}  
with second- and fourth-order terms, where $U$ is the chiral field, $L^\mu = 
U^\dagger \partial^\mu U$, and $f_\pi$ is the pion decay constant.
The relative strength of the fourth-order term is fixed by the parameter $e$. 
In Fig.\ 1 we show the resulting Skyrme form factor for several values of $e$ 
compared with some exponential and monopole curves.

\vskip 1cm 
\hfil
\begin{minipage}{10cm}
\epsfxsize=10cm
\epsfbox{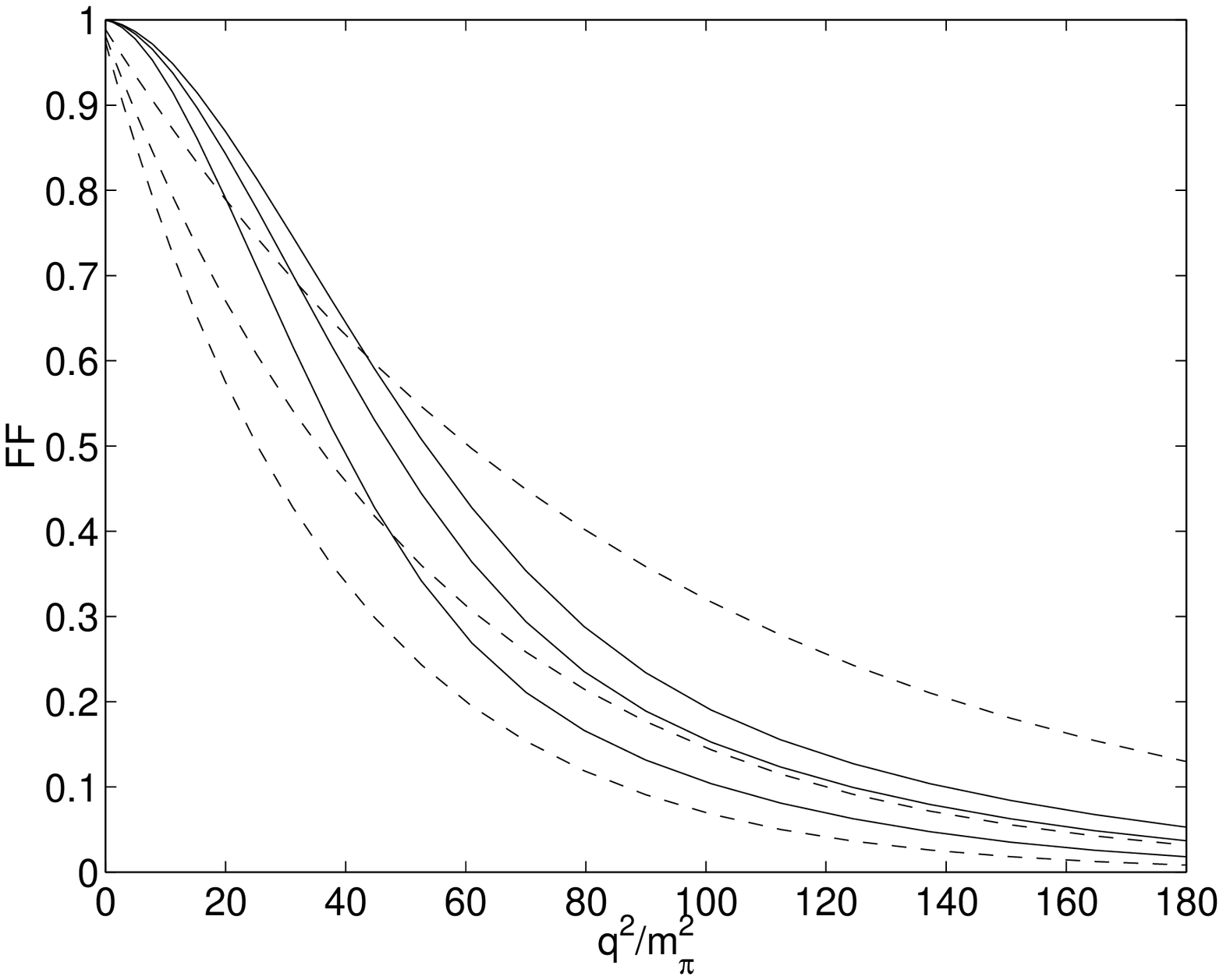} 
\openup -7pt
\it 
FIG. 1. The Skyrme form factor for $e=3.5$, $3.3$ and $3.0$ (solid lines from 
top to bottom) and the exponential form factor for $\Lambda= 1300$, $1000$ 
and $850$ {\rm MeV} (dashed lines from top to bottom) 
\end{minipage} \hfil
\vskip 1cm

This form factor is tested in \cite{holzw96} for the $NN$-system and gives 
reasonable results for the phase shifts of neutron-proton scattering for 
$e=3.5$.
We show the curves also for $e=3.3$ and $e=3.0$ to demonstrate that the 
transition between the hard and the soft part of the form factor tends to 
get sharper when $e$ gets smaller. 
Note that usually values between 3.5 and 4.5 are used to describe baryonic 
observables \cite{holzw96}.
In the following we apply these Skyrme form factors to the meson cloud 
picture of the nucleon. 

\vskip 1cm 
\hfil
\begin{minipage}{8.6cm}
\epsfxsize=7cm
\epsfbox{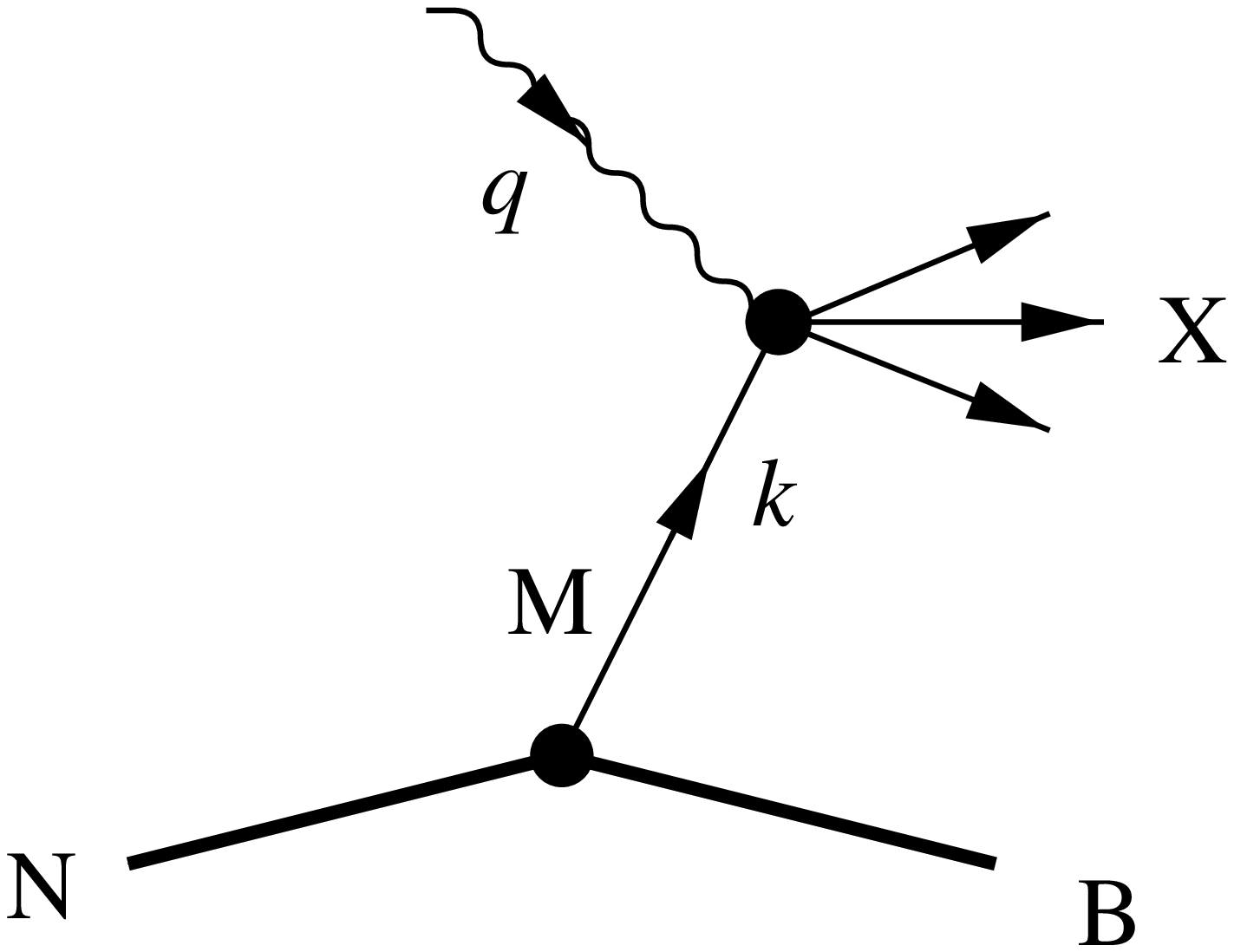} 
\openup -7pt
\it 
FIG. 2. Scattering off the nucleon's meson cloud. The nucleon N splits up in a
baryon $B$ and a meson $M$ with four-momentum $k$. The incoming photon with 
momentum $q$ hits the meson and leaves an unknown final state $X$. 
\end{minipage} \hfil
\vskip 1cm

We consider the process in Fig.2, where a nucleon $N$ splits up in a meson $M$ 
and a baryon $B$ and the meson is hit by the photon. 
From this process corrections to the scattering tensor of the nucleon arise 
which can be translated into corrections $\delta q$ to the quark distributions. This is done by the well known convolution formula \cite{kumano,fra95}
\begin{equation}
  \delta \bar{q}_N (x, Q^2) = \sum_{MB} \int_x^1 f_{MBN}(y) \frac{x}{y} 
  \bar{q}_{M} \left( \frac{x}{y}, Q^2 \right) \, d y
\end{equation}
for the antiquarks $\bar{q}_N$ of a given flavor in the nucleon. 
$\bar{q}_M$ is the antiquark distribution of the same flavor in the meson $M$ 
and the sum runs over all meson-baryon channels taken into account.
$y$ is the momentum fraction carried by the meson in the infinite momentum 
frame and $f_{MBN}(y)$ the momentum distribution of the meson in the nucleon.
For pseudoscalar mesons and spin-$\frac{1}{2}$ fermions as the final baryon 
$B$ one has \cite{fra95}
\begin{equation}
  f_{MBN} = I_{MBN} \frac{g_{MBN}^2}{16\pi^2} y \int_{t_{min}}^\infty 
  \frac{t+(M_B-M_N)^2}{{(t+m_M^2)}^2} G_{MBN}^2(t) \, d t
\end{equation}
and analogously for spin-$\frac{3}{2}$ fermions \cite{fra95} 
\begin{equation}
  f_{MBN} = I_{MBN} \frac{g_{MBN}^2}{16\pi^2} y \int_{t_{min}}^\infty 
  \frac{ {\left( t+(M_B+M_N)^2 \right)}^2 \left( (M_B-M_N)^2 +t \right) }
  { 12M_N^2 M_B^2 {(t+m_M^2)}^2} G_{MBN}^2(t) \, d t,
\end{equation}
where $M_N$, $M_B$, and $m_M$ are the masses of the initial state nucleon, 
the final state baryon, and the meson respectively. Furthermore, $I_{MBN}$ is 
the isospin weight of the $MBN$ vertex and the lower bound of the integration is 
\begin{equation}
  t_{min} = -M_N^2 y + \frac{M_B^2 y}{1-y}.
\end{equation}


\vskip 1cm 
\hfil
\begin{minipage}{14cm}
\epsfxsize=14cm
\epsfbox{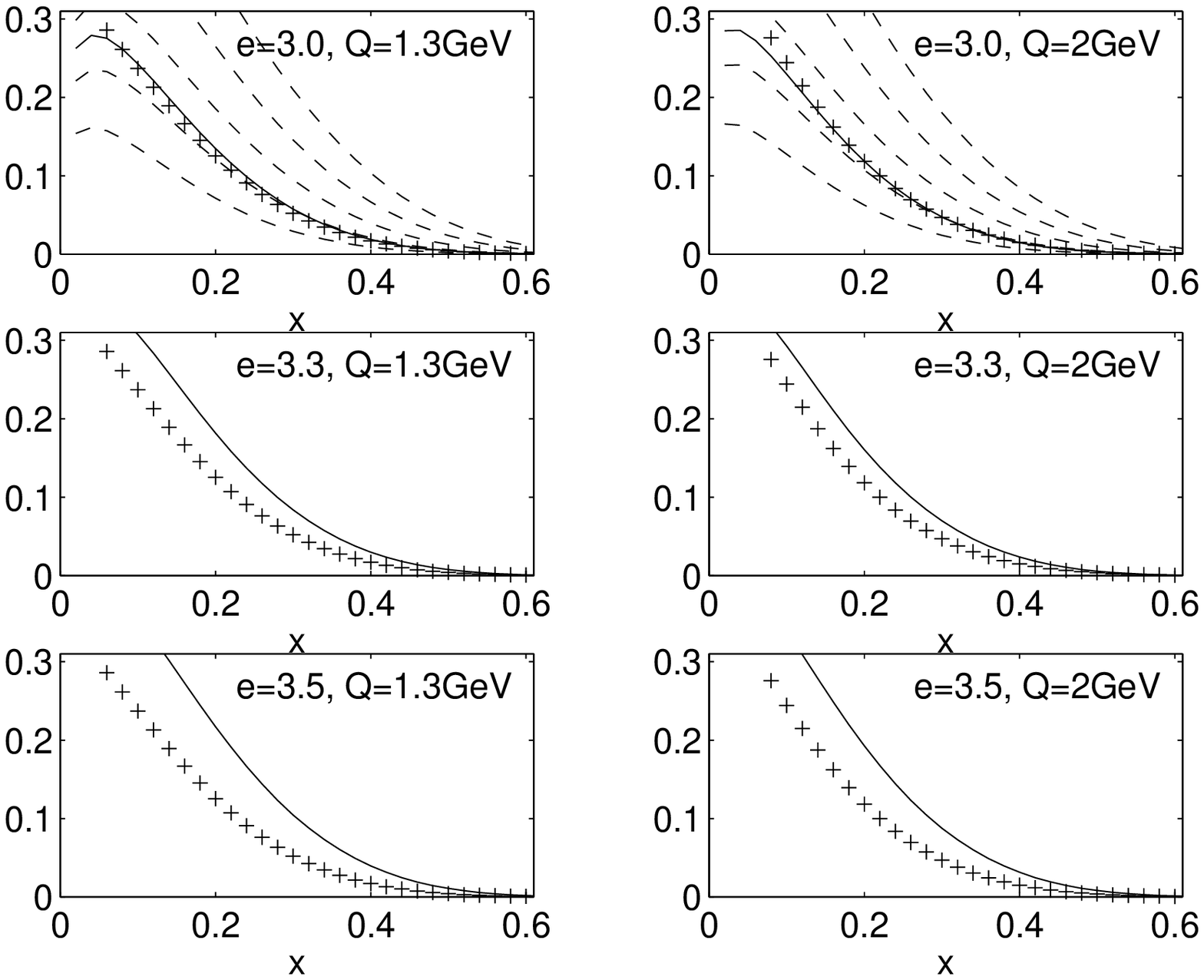} 
\openup -7pt
\it 
FIG. 3. $x\bar{q}_8 = x(\bar{d}+\bar{u})$ for several values of $e$ 
(solid line) and $Q=1.3$ and $2$~{\rm GeV}. The first two plots show in 
addition results for an exponential form factor with cutoff $\Lambda =1450$, 
$1300$, $1150$, $1000$ and $850$~{\rm MeV} respectively (dashed lines from 
top to bottom). The crosses indicate the CTEQ3M fit. 
\end{minipage} \hfil
\vskip 1cm

We calculated the contributions of the $\pi NN$ and $\pi N\Delta$ process to 
$x\bar{u}(x)$ and $x\bar{d}(x)$ using always the Skyrme form factor and the 
pionic parton distributions from \cite{glueck}. 
We compare our results with the the CTEQ3M parton distribution functions 
\cite{cteq3} for $x\bar{q}_3(x) = x(\bar{d}(x) -\bar{u}(x))$ and 
$x\bar{q}_8(x)=x(\bar{d}(x) +\bar{u}(x))$. 
The first quantity is an especially clear signal for meson contributions.
We have no information about the kaon-nucleon form factor from the Skyrme 
model, so we neglect processes containing kaons.

In Figs. 3 and 4 we show our results for $x\bar{q}_3$ and $x\bar{q}_8$ and 
for $Q=1.3~{\rm GeV}$ and $Q=2~{\rm GeV}$ --- with $Q=\sqrt{-q^2}$ measuring 
the photon virtuality and hence the resolution of the DIS process. 
We are also varying the Skyrme parameter $e$. 
For comparison there are also the curves following from an exponential form 
factor with several cutoffs.


\vskip 1cm
\hfil
\begin{minipage}{14cm}
\epsfxsize=14cm
\epsfbox{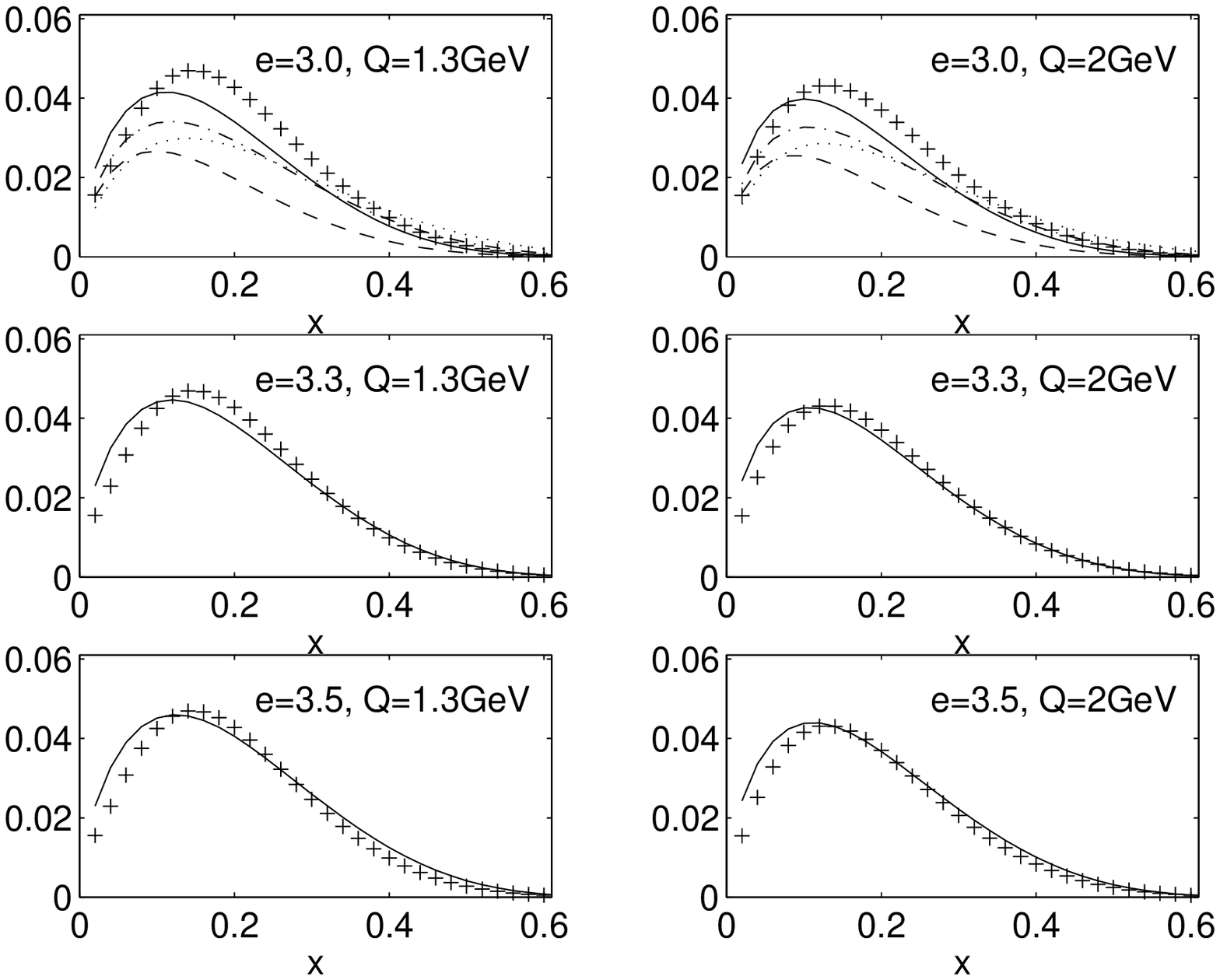} 
\openup -7pt
\it 
FIG. 4. $x\bar{q}_3 = x(\bar{d}-\bar{u})$ for several values of $e$ 
(solid line) and $Q=1.3$ and $2$ {\rm GeV}. The first two plots show in 
addition results for an exponential form factor with cutoff $\Lambda = 850$, 
$1150$ and $1450$ {\rm MeV} (dashed, dashed-dotted, and dotted line). The 
crosses indicate the CTEQ3M data. 
\end{minipage} \hfil
\vskip 1cm

Note also that we act partly at rather high-energy ranges, where the 
applicability of the Skyrme model might be problematic.
Still one can see that $e=3.5$ reproduces well the CTEQ-data for $x\bar{q}_3$. 
For $x\bar{q}_8$ $e=3.5$ leads to an overestimation of the sea and lower 
values of $e$ down to 3.0 are required to fit the data. 
This corresponds to a very strong fourth-order term in the Skyrme Lagrangian 
and is hardly acceptable with view to several low-energy parameters which 
have to be fitted simultaneously with the Skyrme model.

Nevertheless, one can state that the situation is substantially better than in 
the exponential case.
The exponential form factor has the following problems: when the cutoffs for 
the $\pi NN$ and the $\pi N\Delta$ vertex are taken to be the same, 
$x\bar{q}_3$ cannot be fitted. 
Starting at a value of $850~{\rm MeV}$, the contribution increases with 
increasing cutoff, reaching a maximum at approximately $1200~{\rm MeV}$ 
from where on the contribution decreases again for $x<0.5$ without having 
reached the CTEQ-fit.
This comes from the different behavior of the $\pi NN$- and the 
$\pi N\Delta$- sector.
In addition, if one uses cutoffs suggested by low-energy nuclear physics, 
the overestimation of $x\bar{q}_8$ is far worse than for the Skyrme model.

\vskip 1cm
\hfil
\begin{minipage}{8.6cm}
\epsfxsize=8.6cm
\epsfbox{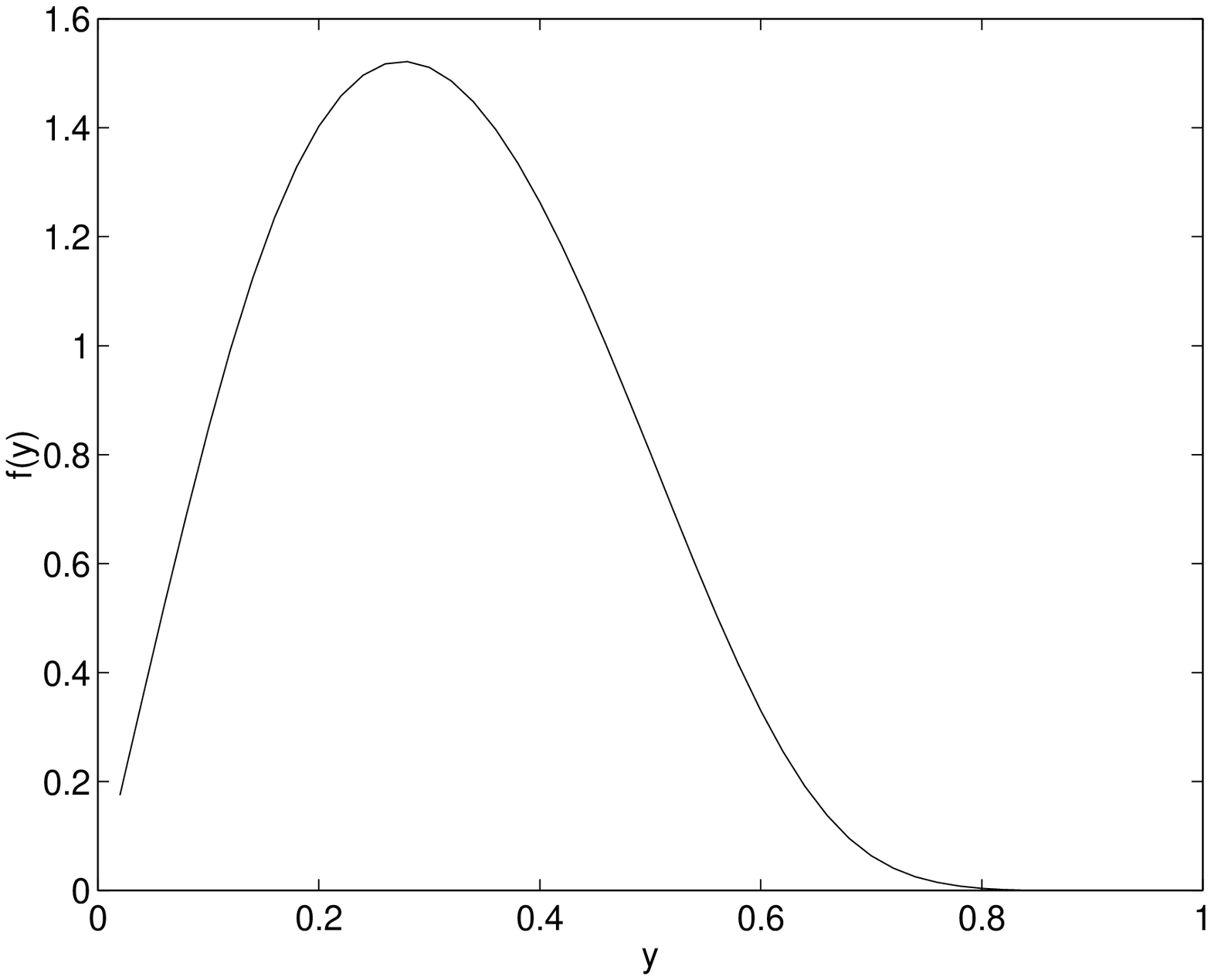} 
\openup -7pt
\it 
FIG. 5. The pion distribution in the nucleon through the $\pi NN$ process for 
a Skyrme form factor with $e=3.5$.
\end{minipage} \hfil
\vskip 1cm

One should also consider the effects of higher order Fock space states. 
For reasonable values of $x$, i.e. $x>0.2$, the convolution integral (6) 
mainly takes into account the tail of the meson distribution $f_{MBN}$ which 
peaks typically around $x \approx 0.25$. 
Fig.5 shows an example for the meson distribution for $e=3.5$.
Multimeson states would obviously blow up the meson distribution for  
$x<0.1$ and reduce it for larger $x$. 
This would lead to a decrease of $x\bar{q}_8(x)$ for $x>0.1$. 
For $x\bar{q}_{3}$ we expect this effect to cancel more or less between the 
$\pi NN$ and the $\pi N\Delta$ contributions.   

We conclude that a form factor which is hard for small $t$ and soft for large 
$t$ is advantageous for meson-cloud models.
We also found that the Skyrme model form factor shows this property naturally.
However, to obtain a good fit to the sea parton distribution, one has to 
choose the Skyrme model parameter $e$ somewhat smaller than permitted by 
fits to low-energy data. We have the hope that higher-order meson states 
tend to reduce the overestimation for $x\bar{q}_8$ without destroying the 
acceptable results for $x\bar{q}_3$ (for $e=3.5$).

\section*{Acknowledgment}

We strongly thank G.\ Holzwarth for making his results available to us prior 
to publication and for numerous helpful discussions. We also thank W.\ Weise 
and L.\ Frankfurt for fruitful discussions.

\end{document}